\documentclass{ifacconf}

\usepackage{amsmath}
\usepackage{amsfonts}
\usepackage{amssymb}
\usepackage{graphicx}
\usepackage{bm}
\usepackage{natbib}
\usepackage{color}
\usepackage[final]{microtype}
\usepackage{float}

\newcommand{\abs}[1]{\left\lvert#1\right\rvert}
\newcommand{\sign}[1]{{\rm{sign}}\left(#1\right)}

\newcommand{\barpow}[1]{\left\lfloor#1\right\rceil}

\begin{document}
\begin{frontmatter}

\title{On the Discretization of Robust Exact Filtering Differentiators} 

\author[First]{J.~E. Carvajal-Rubio} 
\author[Second]{J.~D. S\'anchez-Torres}
\author[Third]{M. Defoort}
\author[First]{A.~G. Loukianov}

\address[First]{\textit{Dept. of Electrical Engineering}, \textit{CINVESTAV-IPN, Guadalajara}, Zapopan, M\'exico, (e-mail: \{jecarvajal,louk\}@gdl.cinvestav.mx)}
\address[Second]{\textit{Dept. of Mathematics and Physics}, \textit{ITESO}, Tlaquepaque, M\'exico, (e-mail: dsanchez@iteso.mx)}
\address[Third]{\textit{LAMIH, CNRS UMR 8201}, \textit{Polytechnic University of Hauts-de-France}, Valenciennes, France, (e-mail:michael.defoort@uphf.fr)}

\begin{abstract}        
This paper deals with the design of discrete-time algorithms for the robust filtering differentiator. Two discrete-time realizations of the filtering differentiator are introduced. The first one, which is based on an exact discretization of the continuous differentiator, is an explicit one, while the second one is an implicit algorithm which enables to remove the numerical chattering phenomenon and to preserve the estimation accuracy properties. Some numerical comparisons between the proposed scheme and an existing discrete-time algorithm show the interest of the proposed implicit discrete-time realization of the filtering differentiator, especially when large sampling periods are considered.
\end{abstract}


\end{frontmatter}

\section{Introduction}

The problems of filtering a noisy signal and differentiation in real-time are crucial issues due to their practical interest in signal processing and control engineering. These problems have been addressed using various methods: Kalman filter \citep{kalman1960new}, algebraic methods \citep{mboup2009numerical}, observation techniques \citep{chitour2002time,spurgeon2008sliding,davila2005second} to name a few. 

Sliding mode techniques are widely used to design observers due to their exceptional accuracy and robustness properties in the presence of matched perturbations \citep{edwards1998sliding,shtessel2014sliding}. However, one of the main disadvantages of these techniques is the chattering phenomenon. High-order sliding mode homogeneous differentiators have been proposed in \citep{Levant_HSMD,levant2011exact}. They give an estimate, in a finite time, of the $n$ derivatives of a signal if its $(n+1)$ order derivative has a known upper bound. Furthermore, they have shown good robustness properties in the presence of noise and exact finite-time convergence in the absence of noise. A filtering differentiator has been investigated in \citep{levant2019robust} in order to exactly differentiate a smooth signal while rejecting a larger class of noises. 

In practice, observation algorithms are usually discretized in order to be implemented in a digital environment. However, the discrete-time approximations of the continuous algorithms are far from being straightforward. Indeed, for high-gain and sliding mode differentiators, an inadequate discrete-time version of the algorithms may lead to numerical chattering \citep{drakunov1990discrete,utkin1994sliding} i.e., high oscillations only due to the numerical methods used in the discretization scheme. 

Several algorithms have been proposed for the implementation of discrete-time sliding mode controllers \citep{drakunov1990discrete,su2000t,nguyen2017improvement,abidi2007discrete}. Concerning the homogeneous differentiator, some explicit discretization algorithms have been derived in \citep{Miki2014,Stefan_dif,koch2019discrete,barbot2020discrete,levant2019robust} in order to preserve the estimation accuracy properties. In \citep{Miki2014}, a discrete-time realization of the homogeneous differentiator, which preserves the computational simplicity of the one-step Euler scheme, has been introduced. In \citep{Stefan_dif}, the proposed discrete algorithm is less sensitive to gain overestimation. A discrete-time differentiator, which includes nonlinear higher-order terms, has been derived in \citep{koch2019discrete} in order to preserve the asymptotic accuracy properties known from the continuous-time differentiator despite the presence of noise. The work in \citep{barbot2020discrete} extends the results from \citep{Miki2014} while also considering non-homogeneous hybrid differentiators. Explicit discrete-time realization of the filtering differentiator has been proposed in \citep{levant2019robust}.

Recently, some implicit discretization algorithms have been investigated in order to ensure a smooth stabilization of the sliding surface in discrete-time for the case without disturbance \citep{acary2011chattering,Brog2019,huber2016implicit,luo2019implicit}. Such algorithms remove the numerical chattering effects due to the time discretization and allow the use of large sampling periods without reducing too much the performances. However, implicit methods have only applied to first-order sliding mode controllers \citep{acary2011chattering}, twisting controllers \citep{huber2016implicit,luo2019implicit} and super-twisting controllers 
\citep{Brog2019}. Nevertheless, an implicit discretization algorithm has been recently proposed in \citep{carvajal2019discretization} for the homogeneous differentiator.

This paper proposes two discretization algorithms, based on the recent results presented in \citep{carvajal2019discretization}, for the robust filtering differentiator given in \citep{levant2019robust}. The first one is an explicit exact discrete-time version of the filtering differentiator, while the second one is an implicit discretization algorithm that removes the numerical chattering effects. Some simulations are given to compare the discrete-time algorithm presented in \citep{levant2019robust} with the proposed ones (explicit and implicit methods). It will be shown that the proposed scheme provides estimates of the derivatives of a given signal with good accuracy and robustness properties even when a large sampling period is considered. 

The rest of the paper is as follows. Section 2 introduces the problem and recalls some preliminaries on the exact filtering differentiator. In Section 3, two discretization algorithms for the robust filtering differentiator are given (i.e., explicit and implicit discrete-time algorithms). At last, in Section 4, some simulations are done to highlight the interest of the proposed scheme when a significant sampling period is considered.

\textbf{Notation.} For $x\in \mathbb{R}$, the absolute value of $x$, denoted by $|x|$, is defined as $|x|=x$ if $x\geq 0$ and $|x|=-x$ if $x<0$. The set-valued function $\sign{x}$ is defined as $\sign{x}=1$ if $x>0$, $\sign{x}=-1$ if $x<0$, and $\sign{x}\in\left[-1,1\right]$ if $x=0$. For $\gamma\geq0$, the signed power $\gamma$ of $x$ is defined as $\barpow{x}^{\gamma}=\left| x\right|^{\gamma}\sign{x}$. In particular, if $\gamma=0$ then $\barpow{x}^\gamma=\sign{x}$. 

\section{Problem statement and preliminaries}

\subsection{Problem statement}

The objective of a differentiator is obtain online the first $n$ derivatives of a function even if there is noise in the measurement. In this paper, this function is represented as $f_0\left( t \right)$, where $f_0:\mathbb{R}\rightarrow \mathbb{R}$. It is also assumed that this function is at least $(n+1)-th$ differentiable and $|f_0^{\left( n+1 \right)}\left( t\right)|\leq L$ for a known real number $L>0$. Furthermore, the input of the differentiator is defined as $f(t)=f_0(t)+\Delta\left(t\right)$. It is also assumed that $\Delta\left( t \right)$ is a Lebesgue-measurable bounded noise with $|\Delta(t)|\leq \delta$ for an unknown real number $\delta>0$. 

In order to compute the derivatives $f_0^{(1)}(t)$, $f_0^{(2)}(t)$, $\cdots$, $f_0^{(n)}(t)$, a state space representation is used. To obtain this representation, the state variables are defined as $x_i(t)=f_0^{(i)}(t)$ and $\bm{x}=\left[ \begin{array}{ccccc}
    x_0 & x_1 & x_2 & \cdots  & x_n
\end{array} \right]^T \in \mathbb{R}^{n+1}$. Therefore, one can obtain the following representation for the differentiation problem in the state space:
\begin{equation}\label{eq:difsys}
\begin{array}{lll}
\dot{\bm{x}}(t)&=&\bm{A}\bm{x}(t)+\bm{e}_{n+1}f_0^{(n+1)}(t)\\
y_o(t)&=&\bm{e}_1^T\bm{x}(t)+\Delta(t)
\end{array}
\end{equation}
with the canonical vectors $\bm{e}_{1}=\left[ \begin{array}{ccccc}
    1 & 0 & \cdots & 0 & 0 
\end{array} \right]^T$, $\bm{e}_{n+1}=\left[ \begin{array}{cccccc}
    0 & 0 & \cdots & 0 & 1 
\end{array} \right]^T$ and $\bm{A}=[\bm{0}_{1\times(n+1)}\;\bm{e}_1\;\bm{e}_2\;\cdots\;\bm{e}_n]$, which is a nilpotent matrix of appropriate dimensions. The representation \eqref{eq:difsys} is interesting in the sense that the successive time derivatives of $f_0\left( t \right)$ can be obtained through the design of a state observer. 

\subsection{Homogeneous high-order differentiator}

In order obtain the first $n$ derivatives of a signal $f_0\left(t\right)$, a continuous-time observer has been proposed in \citep{Levant_HSMD}. For $\Delta(t)=0$, it can be represented in the non-recursive form:
\begin{equation}\label{eq:diflev1}
\dot{\bm{z}}=\bm{A}\bm{z}+\bm{B}\bm{u}\left(\sigma_0\right)   
\end{equation}
where $\bm{u}\left(\sigma_0\right)=\left[\Psi_{0,n}\left( \sigma_0 \right)\; \Psi_{1,n}\left( \sigma_0 \right)\; \cdots \; \Psi_{n,n}\left( \sigma_0 \right) \right]^T$, $\Psi_{i,n}\left( \cdot \right)=-\lambda_{n-i}L^{\frac{i+1}{n+1}} \barpow{\cdot}^{\frac{n-i}{n+1}}$, $\bm{B}$ is identity matrix of appropriate dimensions, $\sigma_0=z_0-x_0$ and $\bm{z}=\left[ \begin{array}{ccccc}
    z_0 & z_1 & z_2 & \ldots  & z_n
\end{array} \right]^T$ is the finite-time estimate of the state vector $\bm{x}$ using adequate parameters $\lambda_i>0$ (see \citep{Reichhartinger_lambda,Levant_dif_fil} for instance). Since the function $ \barpow{z_0-f\left(t\right)}^{0}$ is discontinuous at $z_0=f$, the solutions of system \eqref{eq:diflev1} are understood in the Filippov sense \citep{filippov2013differential}.  

\subsection{Finite-time-exact robust filtering differentiator (FTER)}

Although, differentiator \eqref{eq:diflev1} offers good performance when there exists a Lebesgue-measurable bounded noise $\Delta(t)$ such that $|\Delta(t)|\leq \delta$ with small in average $\delta$, its performance becomes significantly reduced when $\delta$ is large. Due to this reason, in \citet{Levant_dif_fil}, a new finite-time exact robust filtering differentiator has been proposed, with the following structure:
\begin{align}\label{eq:cont_dist_filt}
    \begin{split}
        \dot{\omega}_{i_f}&=-\lambda_{m+1-i_f} L^{\frac{i_f}{m+1}} \barpow{\omega_1}^{\frac{m+1-i_f}{m+1}}+\omega_{i_f+1} \\
         \dot{\omega}_{n_f}&=-\lambda_{n+1} L^{\frac{n_f}{m+1}} \barpow{\omega_1}^{\frac{n+1}{m+1}}+z_{0}-g\left(t\right) \\
        \dot{z}_{i_d}&=-\lambda_{m-i_d} L^{\frac{n_f+1+i_d}{m+1}} \barpow{\omega_1}^{\frac{n-i_d}{m+1}}+z_{i_d+1}\\ 
        i_f&=1, 2, \cdots, n_f-1.  \;\;\;\;\;  i_d=0, 1, 2, \cdots, n.
    \end{split}
\end{align}
where $m=n+n_f$, $n_f\geq 0$, $n_f$ is the filtering order and the parameters $\lambda_i$ are selected as in \eqref{eq:diflev1}. Moreover, $g(t)=f_0\left(t\right)+\upsilon(t)$, where $\upsilon(t)$ is comprised of $n_f+1$ components, $\upsilon(t)=\upsilon_0(t)+\upsilon_1(t)+\cdots+\upsilon_{n_f}(t)$, $\upsilon_j(t)$ is a signal of the global filtering order $j$ and the $jth$-order integral magnitude $\epsilon_j\geq 0$ with $j=0,1,\cdots, n_f$. More details can be founded in \citep{levant2019robust}. In \citep{levant2019robust}, it is shown that differentiator \eqref{eq:cont_dist_filt} offers the following accuracy:
\begin{align*}
    \begin{split}
        |z_i-&f_0^{(i)}\left(t\right)|\leq \mu_i L\rho^{n+1-i},\;\;\mu_i>0,\;\;i=0, 1, 2, \cdots, n.\\
        \rho=&\max \left [ \left(\frac{\epsilon_0}{L} \right)^{\frac{1}{n+1}}, \left(\frac{\epsilon_1}{L} \right)^{\frac{1}{n+2}}, \cdots,  \left(\frac{\epsilon_{n_f}}{L} \right)^{\frac{1}{m+1}}\right]\\
    \end{split}
\end{align*}


\subsection{Discretization (FTER-D)}

In practice, the differentiation algorithms are usually discretized in order to be implemented in a digital environment. In \citep{Levant_dif_fil}, a discrete-time filtering differentiator is presented as follows:
\begin{align}
    \begin{split}\label{eq:FTER-D}
    \begin{bmatrix}
    \bm{w}_{k+1}\\
    \bm{z}_{k+1}
    \end{bmatrix}=
    &\begin{bmatrix}
    \bm{C}\left( \tau \right)_{n_f\times n_f}& \bm{D}\left( \tau \right)_{n_f\times (n+1)} \\ 
    \bm{0}_{(n+1)\times n_f}& \bm{\Phi}(\tau)_{(n+1)\times(n+1)}
    \end{bmatrix}
    \begin{bmatrix}
    \bm{w}_{k}\\
    \bm{z}_{k}
    \end{bmatrix}\\
    &+\tau \bm{e}_{n_f} g_k+\tau \bm{u}_k
    \end{split}
\end{align}
with $\tau=t_{k+1}-t_k$, $\bm{e}_{n_f}=\left[ \begin{array}{cccccccc}
    0 & \cdots & 0 & 1 & 0 & \cdots & 0 
\end{array} \right]^T$, $\bm{w}_k=\bm{w}(\tau k)$, $\bm{z}_k=\bm{z}(\tau k)$, $g_k=g(\tau k)$, $\bm{0}_{(n+1)\times n_f}$ is a matrix whose elements are 0, the matrices $\bm{C}\left( \tau \right)_{n_f\times n_f}$, $\bm{D}\left( \tau \right)_{n_f\times n_f}$ and  $\bm{\Phi}(\tau)_{(n+1)\times(n+1)}$ are defined as:
\small
\begin{align*}
    \begin{split}
     &\bm{C}\left( \tau \right)_{n_f\times n_f}=
    \begin{bmatrix}
    1& \tau & 0 & \cdots & 0 & 0 \\ 
    0& 1 & \tau & \cdots & 0 & 0 \\
    \vdots & \vdots & \vdots & \vdots & \vdots & \vdots \\
    0& 0 & 0 & \cdots & 1 & \tau \\
    0& 0 & 0 & \cdots & 0 & 1 \\
    \end{bmatrix}, \bm{D}\left( \tau \right)_{n_f\times (n+1)}=
    \begin{bmatrix}
    0 & 0 & \cdots & 0 \\ 
    \vdots & \vdots & \vdots & \vdots \\
    0 & 0 & \cdots & 0 \\
    \tau & 0 & \cdots & 0 \\
    \end{bmatrix}\\
    &\bm{\Phi}\left( \tau \right)_{(n+1)\times (n+1)}=
    \begin{bmatrix}
    1 & \tau & \frac{\tau^2}{2!} & \frac{\tau^3}{3!} & \cdots & \frac{\tau^{n-1}}{(n-1)!} & \frac{\tau^{n}}{(n)!} \\ 
    0 & 1 & \tau & \frac{\tau^2}{2!} & \cdots & \frac{\tau^{n-2}}{(n-2)!} & \frac{\tau^{n-1}}{(n-1)!} \\
    \vdots & \vdots & \vdots & \vdots & \vdots & \vdots & \vdots \\
    0 & 0 & 0 & 0 & \cdots & 1 & \tau \\
    0 & 0 & 0 & 0 & \cdots & 0 & 1 \\
    \end{bmatrix}
    \end{split}
\end{align*}
\normalsize

Furthermore $\bm{u}_k$ is defined as:
\begin{align*}
    \begin{split}
    \bm{u}_k&=\left[\Psi_{0,m}\left( \omega_{1,k} \right),\; \cdots, \; \Psi_{m,m}\left( \omega_{1,k} \right) \right]^T
    \end{split}
\end{align*}

For the differentiator \eqref{eq:FTER-D}, $f_{0,k}=f_0(\tau k)$ is assumed as in the continuous differentiator \eqref{eq:cont_dist_filt}, $\upsilon_k=\upsilon(\tau k)$ is comprised of $n_f+1$ components, $\upsilon_k=\upsilon_{0,k}+\upsilon_{1,k}+\ldots+\upsilon_{n_f,k}$, where $\upsilon_{i,k}$ are of the global sampling filtering order $j$ and integral magnitude $\epsilon_{j}$ with $j=0,1,\cdots,n_f$ \citep{levant2019robust}. Furthermore, it is assumed that the set of admissible sampling-time sequences contains sequences for any $\tau>0$. According to \citet{levant2019robust}, the discrete differentiator \eqref{eq:FTER-D} provides the following accuracy: 

\begin{align*}
    \begin{split}
        &|\sigma_{i,k}|\leq \mu_i L\rho^{n+1-i}, \;\;\;\; \mu_i>0, \;\;\; \sigma_{i,k}=z_{i,k}-x_{i,k},\\
        &\rho=\max \left [\tau, \left(\frac{\epsilon_0}{L} \right)^{\frac{1}{n+1}}, \left(\frac{\epsilon_1}{L} \right)^{\frac{1}{n+2}}, \cdots,  \left(\frac{\epsilon_{n_f}}{L} \right)^{\frac{1}{m+1}}\right],\\
        &i=0, 1, 2, \cdots, n. 
    \end{split}
\end{align*}

\section{Discretization of robust exact filtering differentiator}

In this Section, two discrete-time realizations of the filtering differentiator are proposed. The first one is an explicit one, which is based on an exact discretization, while the second one is an implicit algorithm. 

\subsection{Explicit Discretization of the robust exact filtering differentiator (FTER-E)}

Applying the procedure presented in \citep{carvajal2019discretization} to the system \eqref{eq:cont_dist_filt}, the following discrete-time realization of the differentiator is obtained:
\begin{equation}\label{eq:FTER_Exact}
   \begin{bmatrix}
    \bm{w}_{k+1}\\
    \bm{z}_{k+1}
    \end{bmatrix}= \bm{\Phi}\left( \tau \right)_{(m+1)\times (m+1)}
    \begin{bmatrix}
    \bm{w}_{k}\\
    \bm{z}_{k}
    \end{bmatrix}+\bm{h}(\tau) g_k+ \bm{B}^{\ast}(\tau)\bm{u}_k
\end{equation}

Here,
\[
\bm{h}\left( \tau \right)=
    \begin{bmatrix}
    \frac{\tau^{n_f}}{n_f!} \;
    \cdots \;
    \frac{\tau^2}{2!} \;
    \tau \;
    0 \;
    \cdots \;
    0
    \end{bmatrix}^T
\]
and
\[
\bm{B}^\ast\left( \tau \right)=
    \begin{bmatrix}
    \tau & \frac{\tau^2}{2!} & \frac{\tau^3}{3!} & \cdots & \frac{\tau^{m}}{m!} & \frac{\tau^{m+1}}{(m+1)!} \\ 
    0 & \tau & \frac{\tau^2}{2!} &\cdots & \frac{\tau^{m-1}}{(m-1)!} & \frac{\tau^{m}}{m!} \\
    \vdots & \vdots & \vdots & \vdots & \vdots & \vdots \\
    0 & 0 & 0 & \cdots & \tau & \frac{\tau^2}{2!} \\
    0 & 0 & 0 & \cdots & 0 & \tau \\
    \end{bmatrix}
\]

Using Taylor series expansion with Lagrange’s remainders (see \citep{firey1960remainder}) on system \eqref{eq:difsys} the following discrete-time system is obtained:
\begin{equation}\label{eq:disc_lev1}
\bm{x}_{k+1}=\bm{\Phi}\left( \tau \right)_{(n+1)\times (n+1)}\bm{x}_k+\bm{H}_{0,k}    
\end{equation}
with $\bm{H}_{0,k}=\left[ \begin{array}{cccccc}
             \frac{\tau^{n+1}}{\left( n+1 \right)!} f_0^{\left(n+1\right)}\left( \rho_{n} \right)\;
             \cdots
              \tau f_0^{\left(n+1\right)}\left( \rho_0 \right)
\end{array}
\right]^T$,
$\rho_i\in\left(t_k,t_{k+1} \right)$, $\bm{x}_k=\bm{x}(\tau k)$, and $|f_0^{\left(n+1\right)}\left( \rho_i \right)|\leq L$. 

Then, the vector $\left[\bm{w}_{k+1}^T \; \bm{\sigma}_{k+1}^T\right]^T$, with $\bm{\sigma}_k=\bm{\sigma}(\tau k)$ can be represented as:
\begin{align}
    \begin{split}\label{eq:Error_Exact}
    \begin{bmatrix}
    \bm{w}_{k+1}\\
    \bm{\sigma}_{k+1}
    \end{bmatrix}&=\bm{\Phi}(\tau)
    \begin{bmatrix}
    \bm{w}_{k}\\
    \bm{\sigma}_{k}
    \end{bmatrix}+ \bm{B}^{\ast}(\tau)\bm{u}_k -\bm{H}_k+\ldots \\
    &+
    \begin{bmatrix}
    \bm{0}_{n_f\times n_f}& \bm{E}\left( \tau \right)_{n_f\times (n+1)} \\ 
    \bm{0}_{(n+1)\times n_f}& \bm{0}_{(n+1)\times(n+1)}
    \end{bmatrix}
    \begin{bmatrix}
    \bm{0}_{(n_f\times 1)}\\
    \bm{x}_{k}
    \end{bmatrix}
    \end{split}
\end{align}
where $\bm{H}_k=[0\; \;0 \; \;\cdots \; \; 0 \; \; 0 \; \;\bm{H}_{0,k}^T]^T$, the vector of errors is defined as $\bm{\sigma}_k=[\sigma_0(\tau k),\; \sigma_1(\tau k), \; \cdots \; \sigma_n(\tau k)]$ and 
\begin{align*}
    \begin{split}
    \bm{E}\left( \tau \right)_{(n_f \times (n+1))}=
    \begin{bmatrix}
    0& \frac{\tau^{(n_f+1)}}{(n_f+1)!} & \frac{\tau^{(n_f+2)}}{(n_f+2)!} & \cdots & \frac{\tau^{m}}{m!} \\
    \vdots & \vdots & \vdots & \vdots & \vdots \\
    0 & \frac{\tau^{2}}{2!} & \frac{\tau^{3}}{3!} & \cdots & \frac{\tau^{(n+1)}}{(n+1)!} \\
    \end{bmatrix}
    \end{split}
\end{align*}

Due to the non-zero elements of $\bm{E}(\tau)_{n_f\times(n+1)}$, differentiator \eqref{eq:FTER_Exact} does not guarantee convergence for functions with unbounded first $n$ derivatives. Therefore, in order to avoid the last term of the error system \eqref{eq:Error_Exact}, the following discretization is proposed based on the structure of \eqref{eq:FTER_Exact}:
\begin{align}
    \begin{split}\label{eq:FTER-E}
    \begin{bmatrix}
    \bm{w}_{k+1}\\
    \bm{z}_{k+1}
    \end{bmatrix}=
    &\begin{bmatrix}
    \bm{\Phi}\left( \tau \right)_{n_f\times n_f}& \bm{G}\left( \tau \right)_{n_f\times (n+1)} \\ 
    \bm{0}_{(n+1)\times n_f}& \bm{\Phi}(\tau)_{(n+1)\times(n+1)}
    \end{bmatrix}
    \begin{bmatrix}
    \bm{w}_{k}\\
    \bm{z}_{k}
    \end{bmatrix}\\
    &+\bm{h}(\tau) g_k+ \bm{B}^{\ast}(\tau)\bm{u}_k 
    \end{split}
\end{align}
where $\bm{G}\left(\tau\right)_{ n_f \times\left(n+1\right)}$ and $\bm{u}_k$ are defined as:
\begin{align*}
    \begin{split}
    \bm{G}\left( \tau \right)_{(n_f \times (n+1))}=
    \begin{bmatrix}
    \frac{\tau^{n_f}}{n_f!}& 0 & \cdots & 0 & 0 \\
    \vdots & \vdots & \vdots & \vdots & \vdots \\
    \frac{\tau^2}{2!}& 0 & \cdots & 0 & 0 \\
    \tau& 0 & \cdots & 0 & 0 \\
    \end{bmatrix}\;\;\bm{u}_k=\begin{bmatrix}
    \Psi_{0,m}\left( \omega_{1,k} \right)\\
    \Psi_{1,m}\left( \omega_{1,k} \right)\\
    \vdots \\
    \Psi_{m,m}\left( \omega_{1,k} \right)
    \end{bmatrix}
    \end{split}
\end{align*}

For the differentiator \eqref{eq:FTER-E} (i.e., FTER-E),  $\bm{E}(\tau)_{n_f\times(n+1)}=\bm{0}_{n_f\times(n+1)}$ and
\begin{align}
    \begin{split}\label{eq:FTER-E-Error}
    \begin{bmatrix}
    \bm{w}_{k+1}\\
    \bm{\sigma}_{k+1}
    \end{bmatrix}=
    &\begin{bmatrix}
    \bm{\Phi}\left( \tau \right)_{n_f\times n_f}& \bm{G}\left( \tau \right)_{n_f\times (n+1)} \\ 
    \bm{0}_{(n+1)\times n_f}& \bm{\Phi}(\tau)_{(n+1)\times(n+1)}
    \end{bmatrix}
    \begin{bmatrix}
    \bm{w}_{k}\\
    \bm{\sigma}_{k}
    \end{bmatrix}\\
    &+ \bm{B}^{\ast}(\tau)\bm{u}_k-\bm{H}_k  
    \end{split}
\end{align}


\subsection{Implicit Discretization (FTER-I)}

Now, consider the implicit discrete-time algorithm of the robust filtering differentiator. From the differentiator \eqref{eq:FTER-E}, the following algorithm is proposed:
\begin{align}\label{eq:FTER-I-p}
    \begin{split}
    \begin{bmatrix}
    \bm{w}_{k+1}\\
    \bm{z}_{k+1}
    \end{bmatrix}=
    &\begin{bmatrix}
    \bm{\Phi}\left( \tau \right)_{n_f\times n_f}& \bm{G}\left( \tau \right)_{n_f\times (n+1)} \\ 
    \bm{0}_{(n+1)\times n_f}& \bm{\Phi}(\tau)_{(n+1)\times(n+1)}
    \end{bmatrix}
    \begin{bmatrix}
    \bm{w}_{k}\\
    \bm{z}_{k}
    \end{bmatrix}\\
    &+\bm{h}(\tau) g_k+ \bm{B}^{\ast}(\tau)\bm{u}_k \\
    \bm{u}_k=&\left[\Psi_{0,m}\left( \omega_{1,k+1} \right),\; \cdots, \; \Psi_{m,m}\left( \omega_{1,k+1} \right) \right]^T\\
    \Psi_{i,m}&\left( \omega_{1,k+1} \right)\in-\lambda_{m-i}L^{\frac{i+1}{m+1}} \barpow{\omega_{1,k+1}}^{\frac{m-i}{m+1}}
    \end{split}
\end{align}

In order to implement the differentiator \eqref{eq:FTER-I-p}, $\omega_{1,k+1}$ needs to be calculated at time $t=t_k$. Using the difference equation of $\omega_{1,k+1}$, the following inclusion is obtained:
\begin{align}\label{eq:inclu_zoh}
    \begin{split}
        w_{1,k+1} &+ a_{m} \barpow{\omega_{1,k+1}}^{\frac{m}{m+1}}+ \cdots  +a_{1}\barpow{\omega_{1,k+1}}^\frac{1}{m+1} +\ldots\\
        &+ b_k \in -a_{0}\sign{\omega_{1,k+1}}
    \end{split}
\end{align}
where $b_{k}= \frac{\tau^{n_f}}{n_f!} (z_{0,k}-g_k)-\sum_{l=1}^{n_f} \frac{\tau^{(l-1)}}{(l-1)!} w_{l,k}$ and $a_{l}=\frac{\tau^{m-l+1}}{\left( m-l+1 \right)!} \lambda_{l} L^{\frac{m-l+1}{m+1}}$, where $a_i\in\mathbb{R}^+$ and $b_k\in\mathbb{R}$. As in \citep{carvajal2019discretization,Brog2019}, a new support variable is introduced as $\xi_{k+1} \in \sign{\omega_{0,k+1}}$. Using a similar scheme that the one presented in \citep{carvajal2019discretization}, $\omega_{1,k+1}$ and $\xi_{k+1}$ are defined as follows:

\begin{itemize}
    \item \textbf{Case 1}: $b_k > a_{0}$. $\xi_{k+1}=-1$ and $\omega_{1,k+1}=-\left(r_0\right)^{m+1}$, where $r_0$ is the unique positive root of the polynomial:
\end{itemize}
\begin{align}
    \begin{split}
        p\left( r\right)=r^{m+1}+a_{m}r^{m}+\cdots+a_1r+\left(-b_k+a_0 \right)\label{eq:pol_cas1}
    \end{split}
\end{align}

\begin{itemize}
    \item \textbf{Case 2}: $b_k \in [-a_{0},a_{0}]$. $\omega_{1,k,+1}=0$ and $\xi_{k+1}=-\frac{b_k}{a_{0}}$.
\end{itemize}
\begin{itemize}
    \item \textbf{Case 3}: $b_k < -a_{0}$. $\xi_{k+1}=1$ and $\omega_{1,k+1}=r_0^{m+1}$, where $r_0$ is the positive root of the polynomial:
\end{itemize}
\begin{align}
    \begin{split}
        p\left( r\right)=r^{m+1}+a_{m}r^{m}+\cdots+a_1r+\left(b_k+a_0 \right)\label{eq:pol_cas2}
    \end{split}
\end{align}

Furthermore, the pair $\omega_{0,k+1}\in \mathbb{R}$ and $\xi_{k+1}\in\left[-1,1\right]$ is unique for each set of values of $a_l$ and $b_k$. With the new variable $\xi_{k+1}$ the differentiator \eqref{eq:FTER-I-p} is implemented as follows: 

\begin{align}\label{eq:FTER-I}
    \begin{split}
    \begin{bmatrix}
    \bm{w}_{k+1}\\
    \bm{z}_{k+1}
    \end{bmatrix}=
    &\begin{bmatrix}
    \bm{\Phi}\left( \tau \right)_{n_f\times n_f}& \bm{G}\left( \tau \right)_{n_f\times (n+1)} \\ 
    \bm{0}_{(n+1)\times n_f}& \bm{\Phi}(\tau)_{(n+1)\times(n+1)}
    \end{bmatrix}
    \begin{bmatrix}
    \bm{w}_{k}\\
    \bm{z}_{k}
    \end{bmatrix}\\
    &+\bm{h}(\tau) g_k+ \bm{B}^{\ast}(\tau)\bm{v}_k \\
    \bm{v}_k=&\left[\widetilde{\Psi}_{0,m}\left( \omega_{1,k+1} \right),\; \cdots, \; \widetilde{\Psi}_{m,m}\left( \omega_{1,k+1} \right) \right]^T\\
    \widetilde{\Psi}_{i,m}&\left( \omega_{1,k+1} \right)=-\lambda_{m-i}L^{\frac{i+1}{m+1}} \abs{\omega_{1,k+1}}^{\frac{m-i}{m+1}}\xi_{k+1}
    \end{split}
\end{align}
\medskip

\begin{rem}
Since $\xi_{k+1}$ is defined for any value of $\omega_{1,k+1}$ and $\sign{0}\in[-1,1]$,  $\xi_{1,k+1}$ is smoother than the function $\sign{\omega_{1,k+1}}$. 
\end{rem}

\begin{rem}
To implement the differentiator \eqref{eq:FTER-I}, $r_0$ needs to be computed when $b_k \notin \left[-a_0,a_0 \right]$. Hence, a root finding method is needed. Here, the Halley's is used \citep{scavo1995geometry}. 
\end{rem}




\section{Simulation results}

In order to analyze and compare the performance of the differentiators \eqref{eq:FTER-D}, \eqref{eq:FTER-E} and $\eqref{eq:FTER-I}$, two variables will be used: the mean square error of $z_i$ in the time interval $[t_{min},t_{max}]$ (denoted $M_i$) and $Y_i$, which is defined as $Y_i=\max \left\{\left|\sigma_{i,k} \right| \in \mathbb{R}\; | \; 10s \leq t_k \leq t_{max} \right\}$. 
In the following simulations, the filtering differentiator has the following parameters $n=3$, $n_f=2$, $\lambda_0=1.1$, $\lambda_1=6,75$, $\lambda_2=20.26$, $\lambda_3=32.24$, $\lambda_4=23.72$ and $\lambda_5=7$. Notice that the parameters $\lambda_i$ are chosen as in \citep{Levant_dif_fil}. Finally, the initial condition for the differentiator is $\left[ \bm{\omega}_0^T \; \bm{z}_0^T \right]^T=\left[0, 0, 0, 0, 0, 0 \right]$.

\begin{figure}[h!]
	\includegraphics[width=0.46\textwidth]{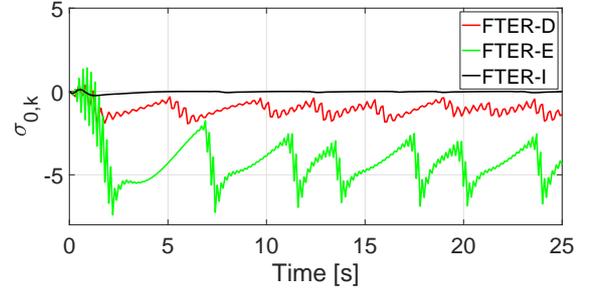}
	\caption{Estimation error for $f_0(t)$.}
	\label{Fig:Function_Estimation_E_1}
\end{figure}

\begin{figure}[h!]
	\includegraphics[width=0.46\textwidth]{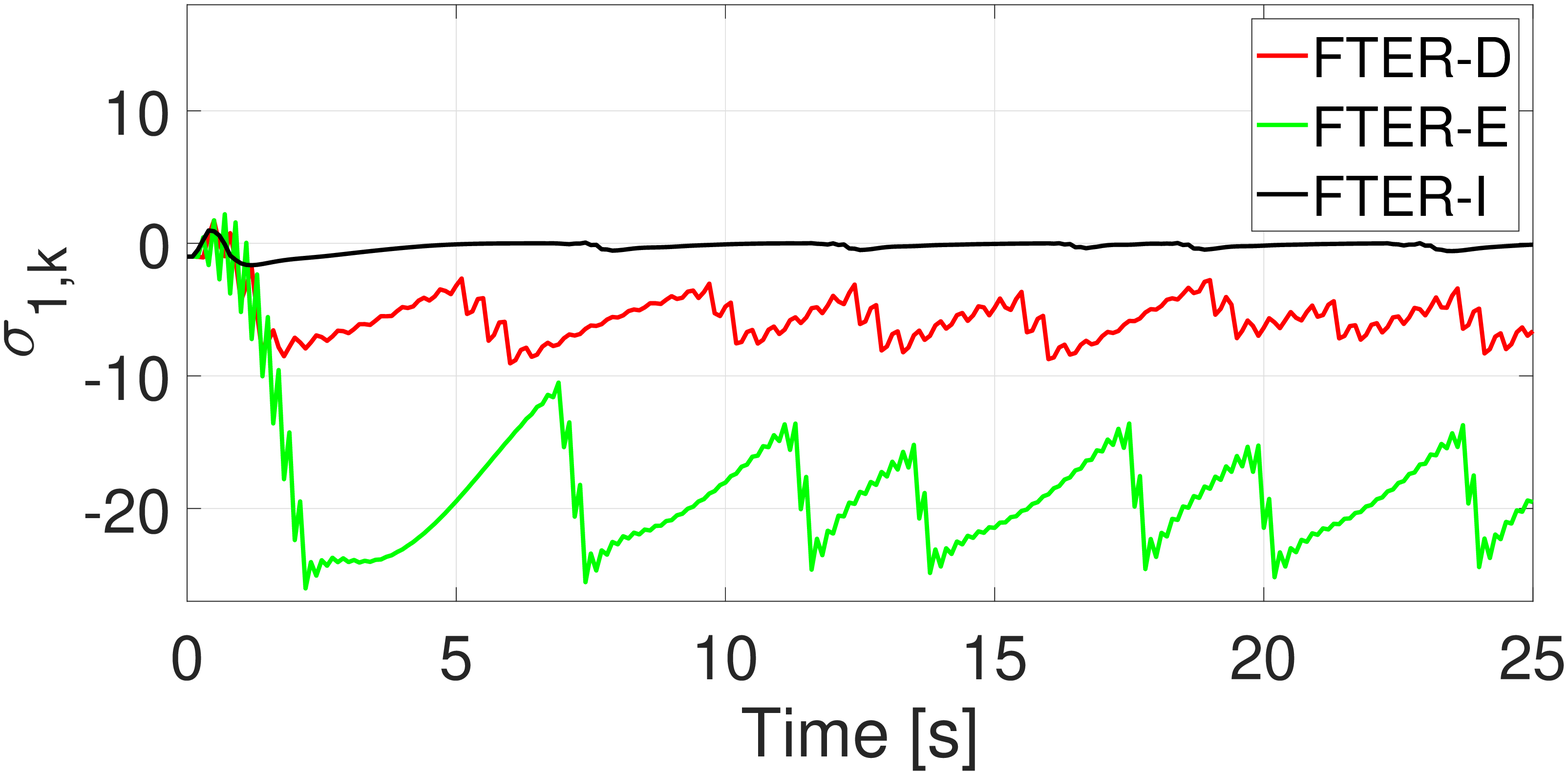}
	\caption{Estimation error for the first derivative of $f_0(t)$.}
	\label{Fig:F_Derivative_Estimation_E_1}
\end{figure}

\begin{figure}[h!]
	\includegraphics[width=0.46\textwidth]{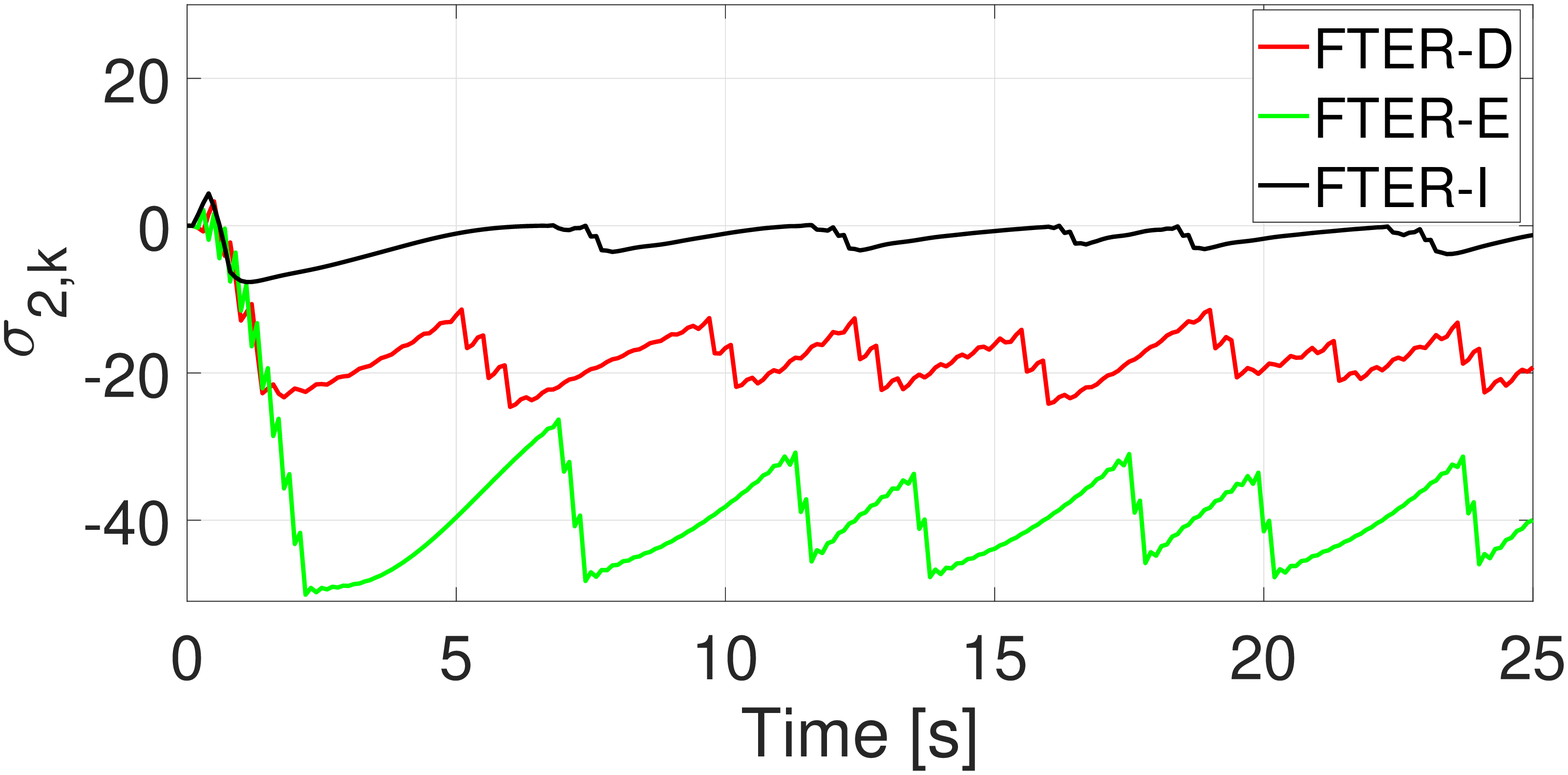}
	\caption{Estimation error for the second derivative of $f_0(t)$.}
	\label{Fig:S_Derivative_Estimation_E_1}
\end{figure}

\begin{figure}[h!]
	\includegraphics[width=0.46\textwidth]{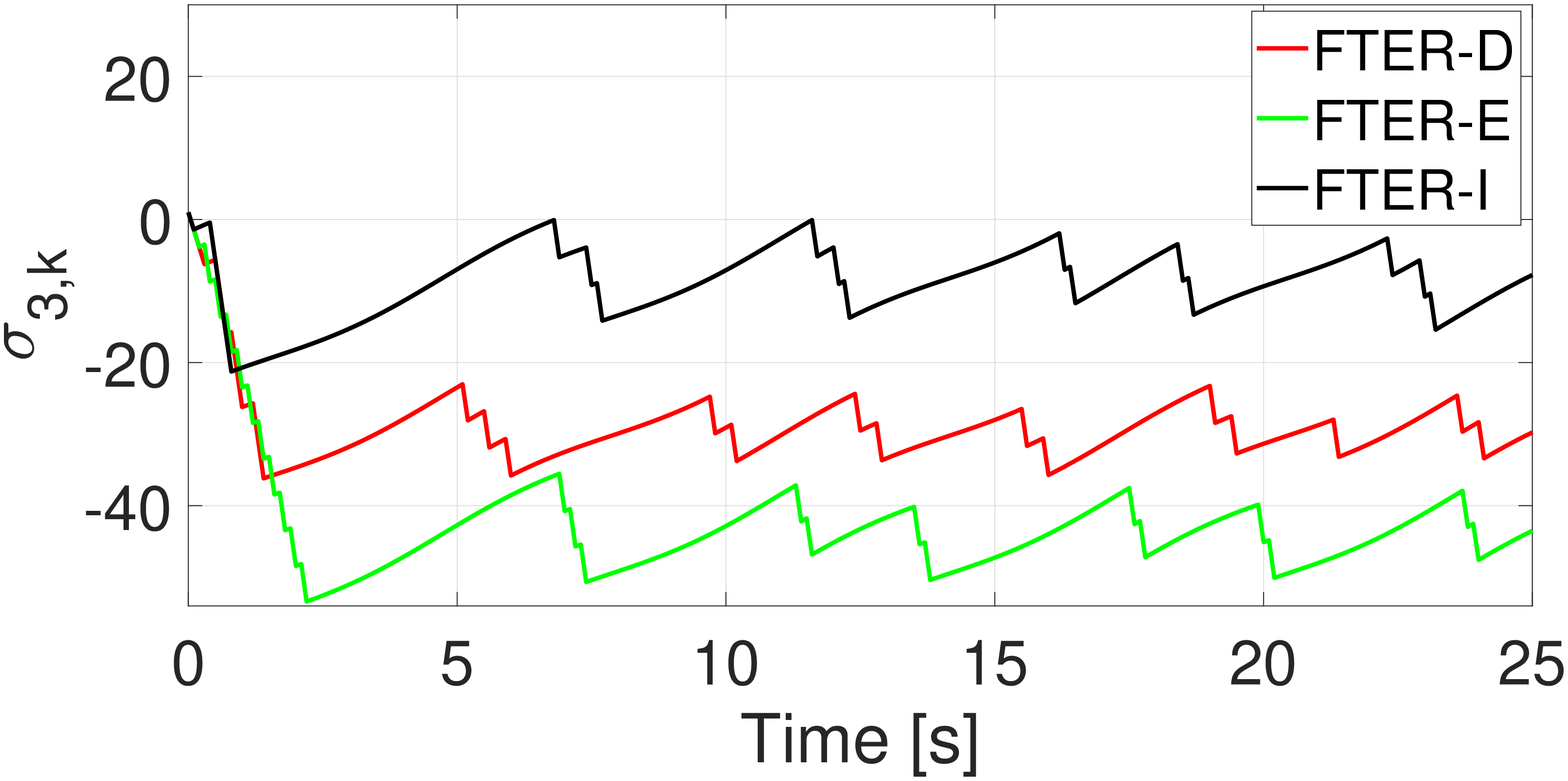}
	\caption{Estimation error for the third derivative of $f_0(t)$.}
	\label{Fig:T_Derivative_Estimation_E_1}
\end{figure}

For the first scenario, $f_0(t)=t^4+\sin(t)$, $L=25$, $\tau=0.1s$, $t_{min}=10s$, $t_{max}=25s$. Furthermore, there is no noise input. Figures \ref{Fig:Function_Estimation_E_1}-\ref{Fig:T_Derivative_Estimation_E_1} show the corresponding estimation errors using the three differentiators. Variables $Y_{i}$ and the $M_i$ are summarised in the Table 1.

\begin{table}[h!]\label{tb:simulation 1}
    \centering
    \begin{tabular}{ |c|c|c|c| } 
 \hline
     & FTER-D & FTER-E & FTER-I  \\ \hline 
 $Y_0$ & 1.8347 & 6.9573 & 0.0736  \\ \hline 
 $Y_1$ & 8.7351 & 25.1874 & 0.5835 \\ \hline
 $Y_2$ & 24.18 & 47.741 & 3.8547 \\ \hline
 $Y_3$ & 35.7077 & 50.3692 & 15.4041 \\ \hline
 $M_0$ & 1.1476 & 4.4312 & 0.0256 \\ \hline
 $M_1$ & 5.9919 & 19.3807 & 0.212 \\ \hline
 $M_2$ & 18.4511 & 39.9016 & 1.7431 \\ \hline
 $M_3$ & 29.7708 & 43.9609 & 8.4592 \\ \hline
\end{tabular}
     \caption{$Y_i$ and $M_i$ for Scenario I.}
\end{table}

For this scenario, the three differentiators converge in finite-time in spite of the unbounded functions $f_0(t)$, $f_0^{(1)}(t)$, $f_0^{(2)}(t)$ and $f_0^{(3)}(t)$. Moreover using the differentiator FTER-I, one obtains the best results as it can been seen in Table 1 and Figures \ref{Fig:Function_Estimation_E_1}-\ref{Fig:T_Derivative_Estimation_E_1}.

\begin{figure}[h!]
	\includegraphics[width=0.46\textwidth]{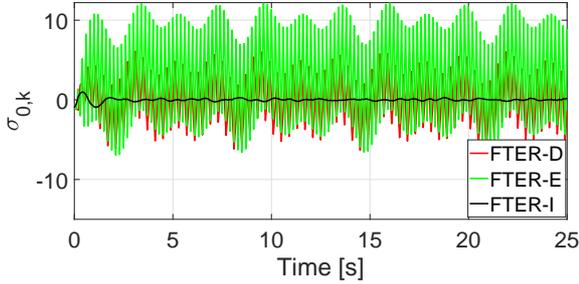}
	\caption{Estimation error for $f_0(t)$.}
	\label{Fig:Function_Estimation_Error_E_2}
\end{figure}

\begin{figure}[h!]
	\includegraphics[width=0.46\textwidth]{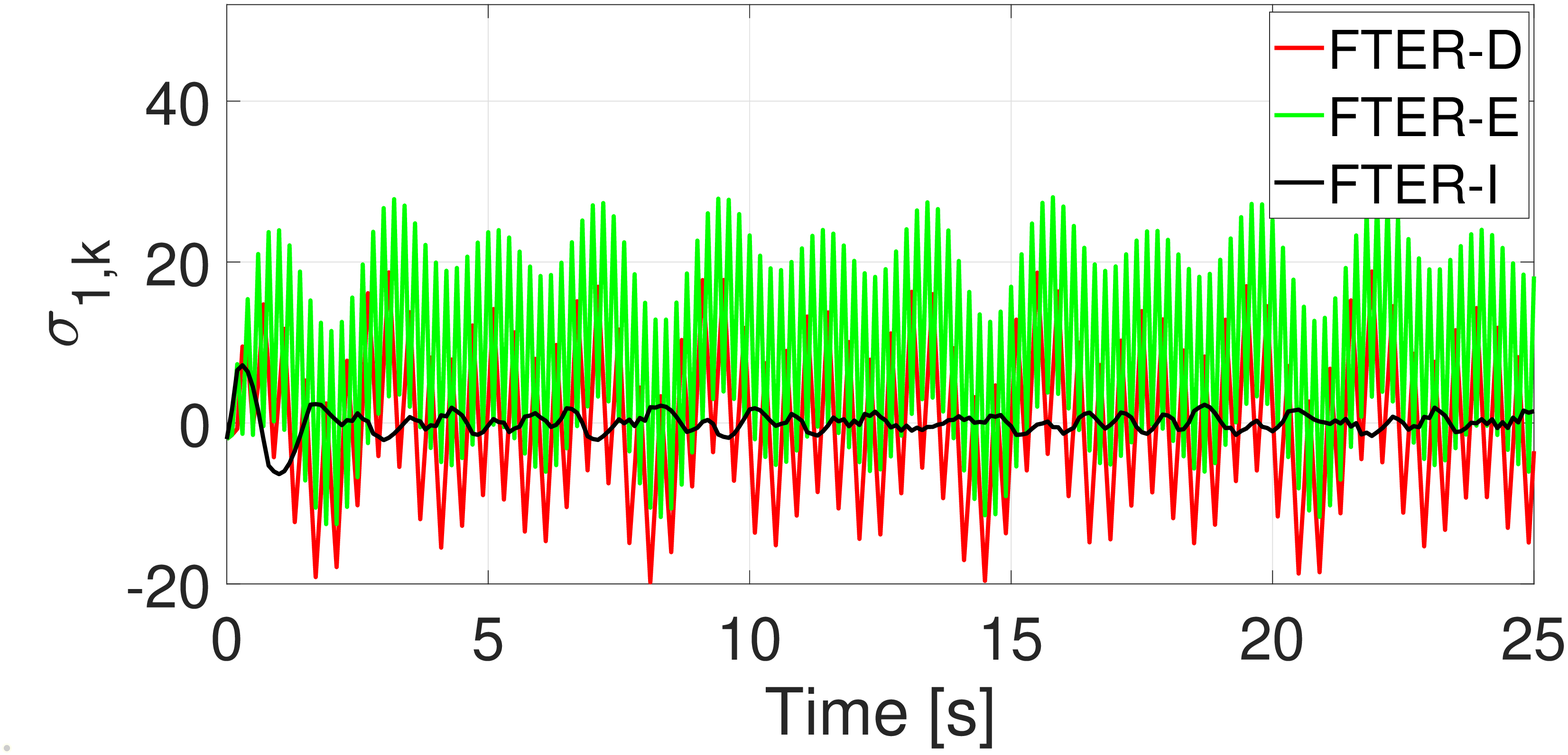}
	\caption{Estimation error for the first derivative of $f_0(t)$.}
	\label{Fig:F_Derivative_Estimation_Error_E_2}
\end{figure}

\begin{figure}[h!]
	\includegraphics[width=0.46\textwidth]{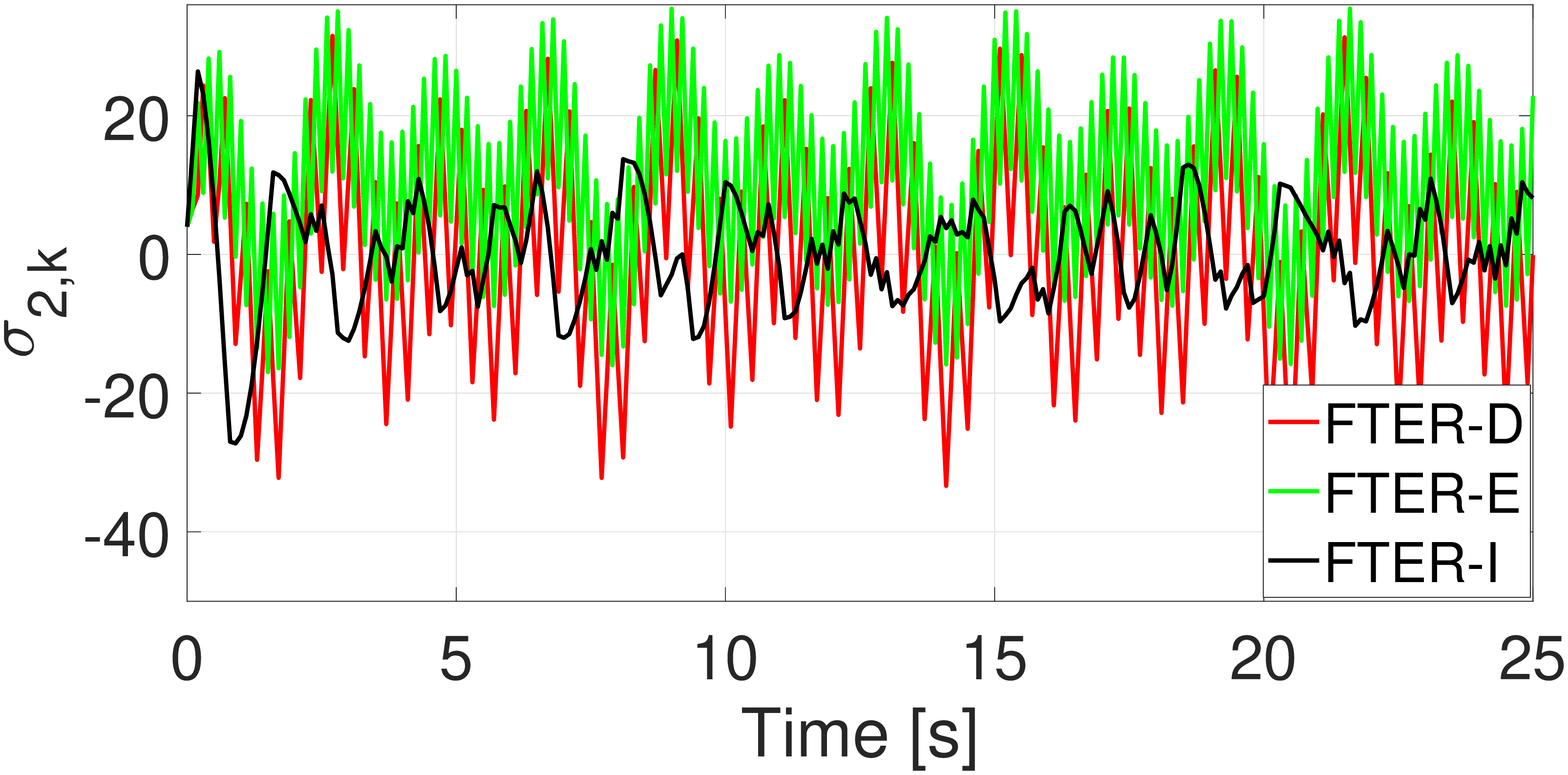}
	\caption{Estimation error for the second derivative of $f_0(t)$.}
	\label{Fig:S_Derivative_Estimation_Error_E_2}
\end{figure}

\begin{figure}[h!]
	\includegraphics[width=0.46\textwidth]{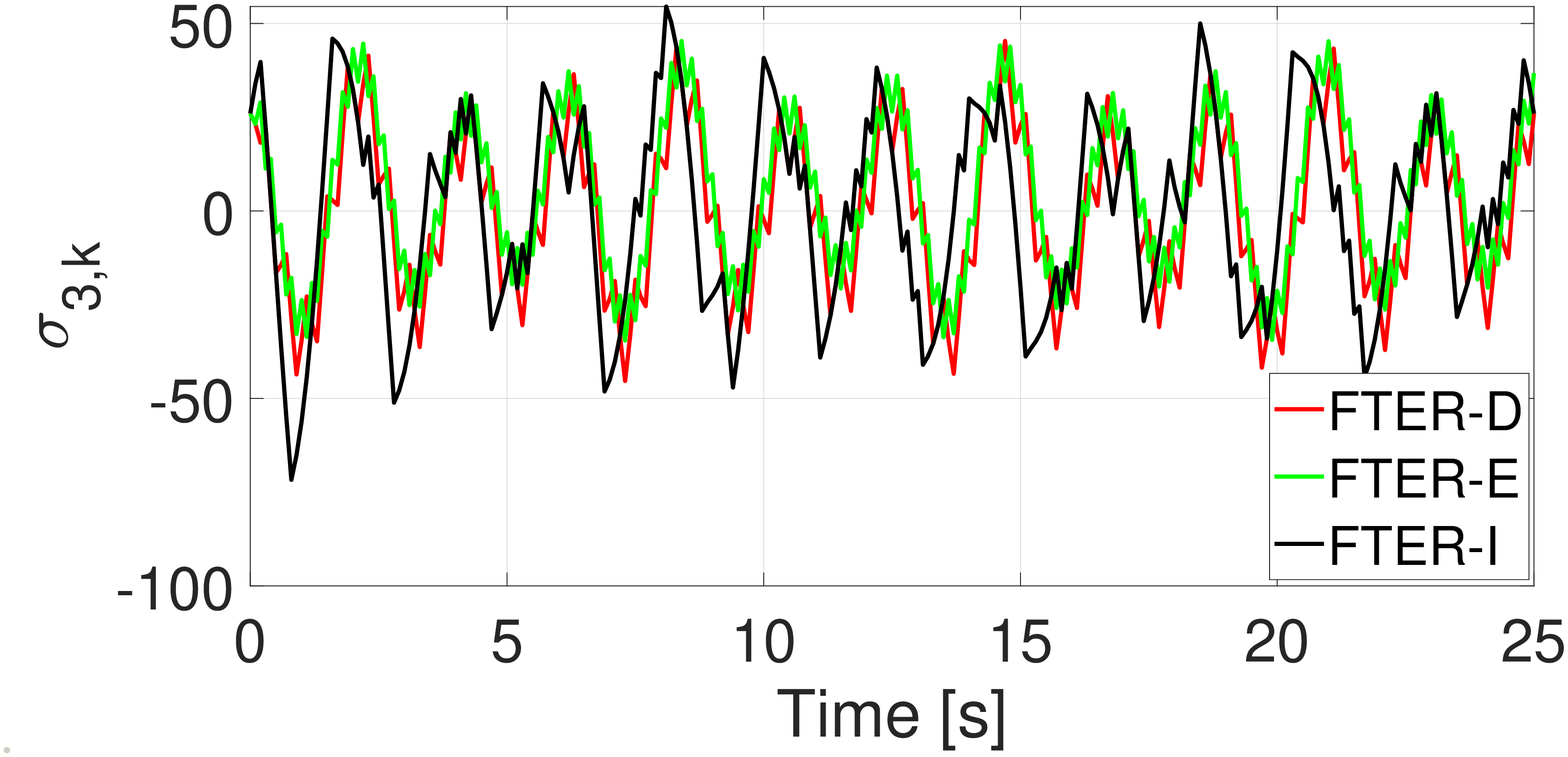}
	\caption{Estimation error for the third derivative of $f_0(t)$.}
	\label{Fig:T_Derivative_Estimation_Error_E_2}
\end{figure}

In the second scenario, $f_0(t)=\sin(3t)+\cos(2t)-\sin(t)+\varepsilon_t$, with $\varepsilon_t\sim\rm{iid}\mathcal{N}(0,0.1^2)$, $L=98$, $\tau=0.1s$, $t_{min}=10s$ and $t_{max}=25s$. Figures \ref{Fig:Function_Estimation_Error_E_2}-\ref{Fig:T_Derivative_Estimation_Error_E_2} show the corresponding estimation errors using the three differentiators. Variables $Y_{i}$ and the $M_i$ are summarised in Table 2.





\begin{table}[h!]\label{tb:simulation 2}
    \centering
    \begin{tabular}{ |c|c|c|c| }
 \hline
     & FTER-D & FTER-E & FTER-I  \\ \hline 
 $Y_0$ & 6.322853 & 12.145781 & 0.274985  \\ \hline 
 $Y_1$ & 19.63017 & 28.059032 & 2.268494 \\ \hline
 $Y_2$ & 33.374147 & 35.453408 & 12.964999 \\ \hline
 $Y_3$ & 45.347208 & 45.279109 & 50.014388 \\ \hline
 $M_0$ & 3.146707 & 7.543292 & 0.097569 \\ \hline
 $M_1$ & 9.009185 & 15.711948 & 0.882423 \\ \hline
 $M_2$ & 14.033387 & 17.492077 & 5.676374 \\ \hline
 $M_3$ & 20.517656 & 20.648051 & 23.703817 \\ \hline
\end{tabular}
    \caption{$Y_i$ and $M_i$ for Scenario II.}
    \label{tab:my_label}
\end{table}

For this scenario, the best result for the first two derivatives has been obtained using the proposed implicit differentiator, i.e., FTER-I. For the last derivative, the explicit differentiators, i.e., FTER-D and FTER-E, present better indexes $Y_3$ and $M_3$ than the implicit one.

\begin{figure}[h!]
	\includegraphics[width=0.45\textwidth]{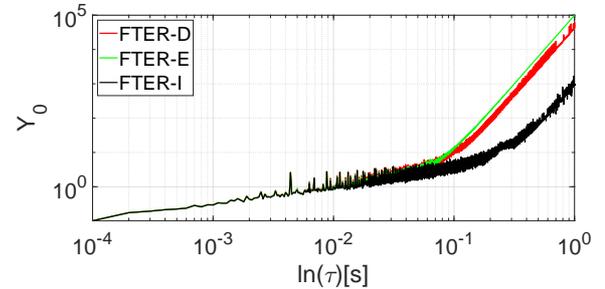}
	\caption{$Y_0$ for $\tau\in\left[0.0001s, 1s\right]$.}
	\label{Fig:Gen_0}
\end{figure}

\begin{figure}[h!]
	\includegraphics[width=0.45\textwidth]{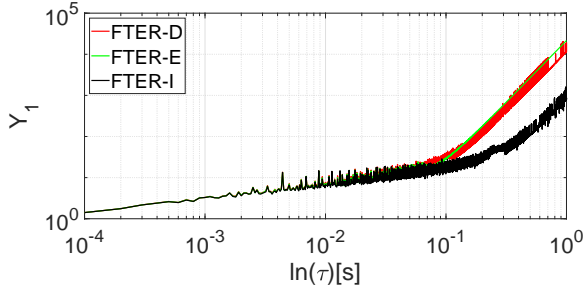}
	\caption{$Y_1$ for $\tau\in\left[0.0001s, 1s\right]$.}
	\label{Fig:F_Gen_1}
\end{figure}

\begin{figure}[h!]
	\includegraphics[width=0.45\textwidth]{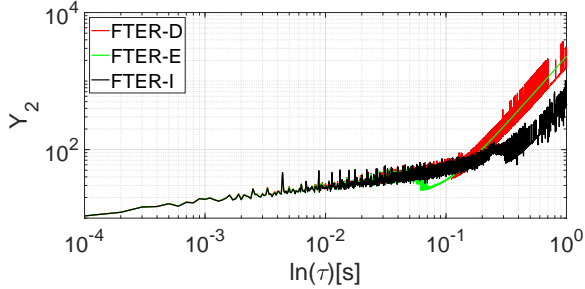}
	\caption{$Y_2$ for $\tau\in\left[0.0001s, 1s\right]$.}
	\label{Fig:Gen_2}
\end{figure}

For the last scenario, in order to test the differentiator under noise and different sampling times, the parameters $Y_i$ are given for different constant sampling times in the interval$\tau\in[0.0001s,1s]$ with a step of $0.0001s$. Furthermore, $f_0(t)=\sin(3t)+\cos(2t)-\sin(t)$, $L=98$, $t_{min}=10s$, $t_{max}=100s$ and the noise is selected as in \citep{levant2019robust}, $\upsilon(t)=\cos(10000t+0.7791)+\varepsilon_t$, with $\varepsilon_t\sim\rm{iid}\mathcal{N}(0,0.5^2)$. The results are summarised in  Figures \ref{Fig:Gen_0}-\ref{Fig:Gen_2}.

\begin{figure}[h!]
	\includegraphics[width=0.45\textwidth]{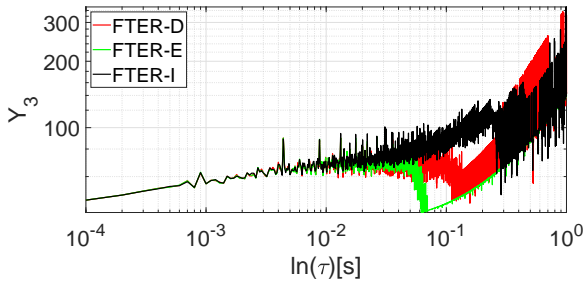}
	\caption{$Y_3$ for $\tau\in\left[0.0001s, 1s\right]$.}
	\label{Fig:Gen_3}
\end{figure}

From Figures \ref{Fig:Gen_0}-\ref{Fig:Gen_2}, one can see that the differentiator FTER-I gives a better performance of the estimation of $f_0(t)$, $f_0^{(1)}(t)$ and $f_0^{(2)}(t)$ or at least similar for the different sampling times. Although, Figures could indicate that for low frequencies, the estimation of the second and third derivatives of the signal is better for the FTER-D and FTER-E compared with FTER-I. 


\section{Conclusion}

Two novel discretization algorithms have been presented for the robust filtering differentiator. The first one, which is based on an exact discretization of the continuous differentiator, is an explicit one, while the second one is an implicit algorithm which enables to remove the numerical chattering phenomenon and to preserve the estimation accuracy properties. Both algorithms have shown a competitive performance in simulations for free-noise input and when the first $n$ derivatives are unbounded. It is also shown a better performance of the current proposal when compared to the discrete version given in \citep{levant2019robust}. Moreover, in simulations and under noise, the FTER-I presents a better estimation for $f_0(t)$, $f_0^{(1)}(t)$, and $f_0^{(2)}(t)$ than the obtained results using FTER-D and FTER-E. Future works will address convergence and robustness proofs for the proposed discretizations.






\bibliography{ifacconf}             

\end{document}